\documentclass[pra,superscriptaddress,floatfix,amsmath,footinbib,amssymb,twocolumn]{revtex4}
\usepackage{amssymb}
\usepackage{amsmath}
\usepackage{epsfig}
\usepackage{t1enc}
\usepackage{soul}
\usepackage{color}
\usepackage{graphicx}
\usepackage{ulem}

\begin{document}

\title{Communication between general-relativistic observers without a shared reference frame}

\author{Agata Ch{\k{e}}ci{\'n}ska}
\author{Andrzej Dragan} 
\affiliation{Institute of Theoretical Physics, University of Warsaw, Pasteura 5, 02-093 Warsaw, Poland}

\date{\today}
\begin{abstract}
We show how to reliably encode quantum information and send it between two arbitrary general-relativistic observers without a shared reference frame. Information stored in a quantum field will inevitably be destroyed by an unknown Bogolyubov transformation relating the observers. However certain quantum correlations between different, independent fields will be preserved, no matter what transformation is applied. We show how to efficiently use these correlations in communication between arbitrary observers. 
\end{abstract}

\maketitle

{\it ---Introduction.}
The central question of the quantum information theory: how to reliably encode, send and decode information \cite{Chuang,Ekert} becomes much more difficult to answer when relativistic effects are taken into account. In the non-relativistic case it is usually implicitly assumed that the sender and receiver share a common reference frame, i.e. they are not moving relative to each other, and their common frame is inertial. As soon as one departs from this assumption one encounters serious conceptual difficulties. It is known that changing an observer's reference frame results in a certain Bogolyubov transformation of the observed state. The most well known consequence of that is the Unruh effect \cite{Unruh}: a vacuum state of a quantum field, as observed by an inertial observer, ceases to be vacuum from the perspective of a uniformly accelerated observer. The latter will detect a thermal state with the temperature proportional to his proper acceleration. Such relativity of the vacuum state is just one example, in general any state will undergo a certain unitary transformation due to motion of the observer. The number of particles, entanglement and other characteristic quantities are affected in general. Furthermore, entering the regime of curved space-times adds more sophistication to the picture, as even the concept of a particle is not well defined and, as a consequence, the notion of a quantum state has no clear interpretation \cite{Birrell}.

In this work we propose a general method of overcoming the problems of mutual communication with quantum states between two observers without a shared reference frame. When one party wishes to send a quantum state to the other, the state becomes distorted due to relative motion. However, following the idea of Ref. \cite{Kok} we note that any type of motion affects states of all quantum fields in an analogous way. Consider a number of independent, non-interacting quantum fields, such as two polarization components of the free electromagnetic field. Although the states of both polarization components will be affected by the relative motion in a certain way, some correlations between the two will remain unaffected. Therefore if the sender and the receiver have access to two or more independent quantum fields, they can securely encode information into correlations between the fields and such information will not be affected by their relative motion. We show how the ability to create and measure these correlations allows the observers to reliably communicate even without sharing a common reference frame. The same method finds application also in more general schemes. For example, this approach can be applied to dynamical space-times that are asymptotically flat, such as the scenario of a collapsing star forming a black hole or an expanding universe modeled by Robertson-Walker space-time \cite{Fabbri,Birrell}. We prove how two observers occupying two asymptotically flat regions of space-time (for example the asymptotic past and the asymptotic future of the expanding universe) can effectively communicate without any knowledge about the details of the intermediate evolution of space-time. This is possible because according to the principle of equivalence, gravity affects all quantum fields in the same way. Therefore certain field correlations will be preserved in the dynamical evolution of the gravitational background.

The idea presented in this work is closely related to the common concept of decoherence-free subspaces used in non-relativistic quantum information to avoid or at least minimize the effect of correlated noise onto communication \cite{Palma, Zanardi, Lidar,Spekkens, Ball2004,Ball2006}; it is also related to the discussion found in \cite{Mancini}. We base our scheme on the observation of \cite{Kok}, where correlations between two components of light polarization were used for communication between two inertial observers without a common reference frame. We generalize this idea to the case of relativistic quantum fields and arbitrary relative types of motion (inertial or not) of the observers related by an unspecified Bogolyubov transformation. Our results can also be applied to other schemes that involve generic Bogolyubov transformations between input and output states of at least two independent quantum fields.

{\it ---The model.} In quantum field theory any change of the coordinate system, for example due to motion of the observer, leads to a certain transformation of all quantum states \cite{Birrell}. In the Heisenberg picture such transformation acting on the field operator under question is linear, since it corresponds to the change of basis of modes between the two coordinate systems. Such a unitary Bogolyubov transformation $\hat{U}$ can always be characterized using a quadratic Hermitian operator $\hat{H}$ via the relation $\hat{U}=\exp\{i\hat{H}\}$. 

Let us consider the simplest example of two bosonic quantum fields of the same type, eg. real scalar and massive fields, $\hat{\phi}$ and $\hat{\tilde{\phi}}$ that are not interacting with each other. An algebra describing arbitrary quadratic Hermitian operators has the following set of generators for $\hat{\phi}$ \cite{Zhang, Luis}:
\begin{align}
\label{generators}
\hat{G}^1_{ij} &= \hat{a}^\dagger_i\hat{a}_j+\hat{a}^\dagger_j\hat{a}_i ,~~~
\hat{G}^2_{ij} = i\left(\hat{a}^\dagger_i\hat{a}_j-\hat{a}^\dagger_j\hat{a}_i\right), \\ 
\hat{G}^3_{ij} &=  \hat{a}_i\hat{a}_j+\hat{a}^\dagger_i\hat{a}^\dagger_j,~~~ 
\hat{G}^4_{ij} = i\left(\hat{a}_i\hat{a}_j-\hat{a}^\dagger_i\hat{a}^\dagger_j\right),\nonumber
\end{align} 
where $\hat{a}_i$ are the annihilation operators corresponding to the decomposition of the field operator $\hat{\phi}$ in the first basis of modes. We have an analogous set of generators $\hat{\tilde{G}}^\xi_{ij}$ for the other field $\hat{\tilde{\phi}}$. Consequently, the generator of an arbitrary Bogolyubov transformation of two non-interacting fields $\hat{\phi}$ and $\hat{\tilde{\phi}}$ of the same type takes the form:
\begin{align}
\label{genham}
\hat{H}=D^\xi_{ij}\hat{G}^\xi_{ij} \otimes \openone + \tilde{D}^\xi_{ij}\openone\otimes\hat{\tilde{G}}^\xi_{ij},
\end{align}
where $\xi\in\{1,2,3,4\}$, $D^\xi_{ij}$, $\tilde{D}_{ij}^\xi$ are arbitrary real coefficients characterizing the Bogolyubov transformation under question, and we use the standard summation convention. Indices $\{i,j\}$ can take either discrete or continuous values depending on the character of the field or chosen boundary conditions. In order to introduce the full symmetry between both fields we take both fields to have equal masses: $m=\tilde{m}$. One physical example of two such fields (when m=0) is two polarization components of the electromagnetic field. In the considered case the generator $\hat{H}$ becomes fully symmetric with respect to the interchange $\hat{\phi}\leftrightarrow\hat{\tilde{\phi}}$ and \eqref{genham} can be written in a simplified form with $D_{ij}^\xi=\tilde{D}_{ij}^\xi$:
\begin{align}
\label{hamiltonian}
\hat{H}=D^\xi_{ij} \left( \hat{G}^\xi_{ij}\otimes \openone + \openone\otimes\hat{\tilde{G}}^\xi_{ij} \right).
\end{align}

The transformation $\hat{U}=\exp\{i\hat{H}\}$ with $\hat{H}$ given by \eqref{hamiltonian} is a general operation acting symmetrically on fields $\hat{\phi}$ and $\hat{\tilde{\phi}}$. The unknown coefficients $D^\xi_{ij}$ in $\eqref{hamiltonian}$ are related to the unknown relative motion between the sender and the receiver. Let us try to use the fields' interchange symmetry to allow the two partners to communicate.

Suppose that the sender and the receiver choose an observable $\hat{L}$ with a discrete spectrum $\lambda_i$ and the sender chooses to encode and send one of the values $\lambda_i$ belonging to that spectrum. She does it by sending to the receiver the eigenstate corresponding to the chosen eigenvalue. In order to retrieve the transmitted information the receiver measures the acquired state using $\hat{L}$. Since the sender and the receiver are in the unknown relative motion, the transmitted eigenstate undergoes some unknown operation $\hat{U}=\exp\{i\hat{H}\}$. In the Heisenberg picture this transformation corresponds to the transformation of the considered observable $\hat{L}\to \hat{U}^\dagger\hat{L}\hat{U}$. We ask: under what circumstances the receiver will be able to retrieve the encoded classical number $\lambda_i$ with his measurement of the observable $\hat{L}$?

Let us notice that if $\hat{L}$ is such that it commutes with the Hermitian operator $\hat{H}$ for an arbitrary choice of the parameters appearing in the equation \eqref{hamiltonian} it will also commute with $\hat{U}=\exp\{i\hat{H}\}$. Consequently the result of the measurement performed by the receiver will inevitably yield the desired eigenvalue $\lambda_i$. It turns out that due to the field interchange symmetry present in \eqref{hamiltonian} there always exists such an operator.

Consider the following observable:   
\begin{align}
\hat{L}&=\hat{x}_k\otimes\hat{\tilde{p}}_k-\hat{p}_k\otimes\hat{\tilde{x}}_k,
\end{align}
where $\hat{x}_k = (\hat{a}_k + \hat{a}_k^\dagger)/\sqrt{2}$, $\hat{p}_k = (\hat{a}_k - \hat{a}_k^\dagger)/\sqrt{2}i$ are quadratures corresponding to the $k$-th mode of the field $\hat{\phi}$ and analogously for the tiled operators. Again, we have used the standard summation convention. To show the invariance of the operator $\hat{L}$ let us write it down in the form:
\begin{align}
\hat{L}=-i\left(\hat{a}_k^\dagger\otimes\hat{\tilde{a}}_k-\hat{a}_k\otimes\hat{\tilde{a}}^\dagger_k\right)\label{L}.
\end{align}
Then it is straightforward to verify explicitly that for all $\xi$ we have $\left[\hat{L},\hat{G}^\xi_{ij}\otimes \mathbb{I} + \mathbb{I}\otimes\hat{\tilde{G}}^\xi_{ij}\right] = 0$. Using the equations \eqref{generators} and \eqref{L} we can write down explicitly one of the commutators involved in this calculation. Choosing $\xi=1$ (again, the summation convention holds) we have:
\begin{align}
&\left[\hat{L},D^1_{ij} \left( \hat{G}^1_{ij}\otimes \openone + \openone\otimes\hat{\tilde{G}}^1_{ij} \right)\right]=\nonumber\\
&-iD_{ij}^1\bigg\{-\delta_{kj}\hat{a}_i^\dagger\hat{\tilde{a}}_k-\delta_{ki}\hat{a}_j^\dagger\hat{\tilde{a}}_k+\delta_{ki}\hat{a}_k^\dagger\hat{\tilde{a}}_j+\delta_{kj}\hat{a}_k^\dagger\hat{\tilde{a}}_i+\nonumber\\
&-\delta_{ki}\hat{a}_j\hat{\tilde{a}}^\dagger_k-\delta_{kj}\hat{a}_i\hat{\tilde{a}}^\dagger_k+\delta_{kj}\hat{a}_k\hat{\tilde{a}}^\dagger_i+\delta_{ki}\hat{a}_k\hat{\tilde{a}}^\dagger_j\bigg\}\nonumber\\
&=-iD_{ij}^1\big\{0\big\}=\,0.
\end{align}
The rest of the commutators can be evaluated in the same manner. As a consequence, we obtain:
\begin{align}
\left[\hat{L},\hat{H}\right] = 0.\label{commutator}
\end{align}
The above equation shows that the operator $\hat{L}$ commutes with the considered Bogolyubov transformation and therefore it is an appropriate observable for encoding information into a pair of quantum fields. The transmitted information remains robust against the influence of the relative motion of the observers. Let us also notice that the eigenstates of the operator $\hat{L}$ used to encode information involve entanglement of the two considered fields therefore both the sender and receiver must be capable of preparing and measuring such entangled states. The eigenstates of the operator $\hat{L}$ may in general change, however the eigenspectrum of the operator will remain the same when measured in different reference frames.


{\it ---Eigenstates.} Let us determine the eigenstates of the $\hat{L}$ operator in the position (quadrature) representation. We first define:
\begin{align}
f_{\lambda,k}(x_k,\tilde{x}_k)&=e^{i\lambda\arctan(x_k / \tilde{x}_k)},\label{eigenfunction_1} ~~~\lambda\in\mathbb{N}
\end{align}
which is an eigenstate of the operator $\hat{x}_k\hat{\tilde{p}}_k-\hat{p}_k\hat{\tilde{x}}_k$ for fixed $k$. This can be seen by noting that the generalized angular momentum operator $\hat{x}_k\hat{\tilde{p}}_k-\hat{p}_k\hat{\tilde{x}}_k$ has eigenstates given by phase factors $e^{i \lambda\varphi}$, where $\lambda\in\mathbb{N}$ and $\varphi$ is a generalized angle, in this case $\varphi = \arctan(x_k / \tilde{x}_k)$. Function $f_{\lambda,k}(x_k,\tilde{x}_k)$ is unnormalized. We note however that it still remains an eigenfunction of $\hat{x}_k\hat{\tilde{p}}_k-\hat{p}_k\hat{\tilde{x}}_k$ after multiplication by an arbitrary (normalized) function of $(x_k^2+\tilde{x}_k^2)$. Therefore, an arbitrary eigenfunction of the operator $\hat{L}$ has the following form:
\begin{align}
\label{eigenfunction_2}
F_{\lambda}(\{x,\tilde{x}\})&=\Pi_k f_{\lambda,k}(x_k,\tilde{x}_k) g_{k}(x^2_k+\tilde{x}^2_k),
\end{align}
where $g_k$ are arbitrary, normalizable functions, for example Gaussians: $g_{k}(x)\sim\exp(-x^2)$. As noted above, the eigenvalue $\lambda$ belongs to the discrete set of natural numbers $\mathbb{N}$: $\hat{L}F_{\lambda} = \lambda F_{\lambda}$. Similar states were discussed in \cite{Kok} alongside the scenario involving a less general case of Lorentz transformations between inertial reference frames. Due to Eq.~\eqref{commutator}, the spectrum of eigenvalues $\lambda$ is invariant under operation $\hat{U}$ and can be retrieved after the transformation by measuring the observable $\hat{L}$. One has to note however that for the operator $\hat{L}$ involving all the modes $k$, one needs to introduce additional normalization. Note that one has $\hat{L}F_{\lambda}=\int \text{d}k \hat{L}_kF_{\lambda}=\lambda\big[\int \text{d}k \,1\big] F_{\lambda}$. Therefore, the definition of $\hat{L}$ needs to be equipped with a normalization function $\rho(k)$, $\int \text{d}k \rho(k)=1$, so that we obtain 
\begin{align}
\hat{L} F_{\lambda}=\int \text{d}k \rho(k)\hat{L}_k F_{\lambda}=\lambda\int \text{d}k \,\rho(k) F_{\lambda}=\lambda F_{\lambda}.\label{L-normalization}
\end{align}
Our construction is very general as it can be applied to various systems in which Bogolyubov transformations play a role. Here we work in the $(1+1)$ dimensional case, however our formalism can be immediately applied to the $(3+1)$ dimensional one. In the following, we consider two examples in which one can see explicitly how the invariant operator can be used for communication purposes.

{\it ---Example a: expanding universe.} Consider the case of the expanding universe described by a two-dimensional Robertson-Walker model characterized by a metric:
\begin{align}
\label{dS}
\text{d}s^2=C(\tau)(\text{d}\tau^2-\text{d}x^2),\;C(\tau)=1+\epsilon(1+\tanh\sigma\tau),
\end{align}
with $\{\epsilon,\sigma\}\in\mathbb{R}^+$. Suppose that an observer in the distant past wishes to encode an integer number into the quantum state of the field and send it over to the observer that will receive it in the asymptotic future. We assume that they lack the detailed knowledge about the spacetime expansion. To be strict, let us assume that the sender and the receiver do not know the expansion rate $\sigma$ and its magnitude $\epsilon$. The asymptotic past and the future of the metric \eqref{dS} are conformally equivalent to Minkowski spacetime, therefore the definition of quantum states in these regions exists and our problem is well defined. Let us take two identical scalar real and massive fields $\hat{\phi}$ and $\tilde{\hat{\phi}}$ existing in the expanding universe and study the solutions of the corresponding Klein-Gordon equation in the asymptotic regions:
\begin{align}
(\square+m^2)\hat{\phi}(x)=0,\label{KG}
\end{align}
and similarly for $\tilde{\hat{\phi}}$. The full analysis of the solutions to this equation can be found in \cite{Birrell}. The asymptotic solutions in the past and in the future, respectively, take the following form:
\begin{align}
\bar{u}_k(\tau,x)&\longrightarrow_{\tau\to -\infty} (4\pi\bar{\omega}_{k})^{-1/2}e^{i(kx-\bar{\omega}_{k}\tau)},\nonumber\\
u_k(\tau,x)&\longrightarrow_{\tau\to +\infty} (4\pi\omega_k)^{-1/2}e^{i(kx-\omega_k\tau)},
\end{align}
where $\bar{\omega}_k=[k^2+m^2]^{1/2}$ and $\omega_k=[k^2+m^2(1+2\epsilon)]^{1/2}$. Let us denote the corresponding annihilation operators in the past with $\hat{a}_k$ and in the future with $\hat{b}_k$ then the Bogolyubov transformation between the two has a very simple block-diagonal form \cite{Birrell} (from now on we suppress the summation over $k$):
\begin{align}
\hat{b}_k&=\alpha_k^*\hat{a}_k-\beta_k\hat{a}_{-k}^\dagger,\nonumber\\
\hat{b}_{-k}&=\alpha_{-k}^*\hat{a}_{-k}-\beta_{-k}\hat{a}_{k}^\dagger,\label{expanding_Bogo}
\end{align}
with an analogous transformation for modes of the field $\tilde{\phi}$ (the explicit form of coefficients $\alpha_k$ and $\beta_k$ that can be found in \cite{Birrell}; they can be always made real by absorbing their complex phases into re-defined annihilation operators). Here, and from now on, we suppress the summation convention. Without a loss of generality, we can limit ourselves to analyzing the Hilbert subspace spanned by the wavevectors $\{k,-k\}$ and work effectively with four-dimensional Hilbert space of two fields. Consequently, we can consider the following operator that generates the Bogolyubov transformation:
\begin{align}
\hat{H}_k=i\left(\xi_k^*\hat{a}_k\hat{a}_{-k}-\xi_k \hat{a}_k^\dagger \hat{a}_{-k}^\dagger+\xi_k^*\hat{\tilde{a}}_{k}\hat{\tilde{a}}_{-k}-\xi_k\hat{\tilde{a}}_k^\dagger \hat{\tilde{a}}_{-k}^\dagger\right),
\end{align}
where $\xi_k$ characterizes the details of expansion. We introduce the corresponding invariant operator $\hat{L}_k$, such that $[\hat{L}_k,\hat{H}_k]=0$ 
\begin{align}
\hat{L}_k&=\hat{x}_k\hat{\tilde{p}}_k-\hat{p}_k\hat{\tilde{x}}_k + \hat{x}_{-k}\hat{\tilde{p}}_{-k}-\hat{p}_{-k}\hat{\tilde{x}}_{-k}.
\end{align} 
Its eigenstates can be easily written down based on the discussion presented in the previous paragraphs. For the $k$-th sector we have
\begin{align}
F_{\lambda,k}(x_k,x_{-k},\tilde{x}_k,\tilde{x}_{-k})&=f_{\lambda,k}(x_k,\tilde{x}_k) g_{k}(x^2_k+\tilde{x}^2_k)\times\\
\times &f_{\lambda,-k}(x_{-k},\tilde{x}_{-k}) g_{-k}(x^2_{-k}+\tilde{x}^2_{-k}).\nonumber
\end{align}
The above four-mode eigenstates can be used by the observer in the distant past to reliably encode and send a natural number $\lambda$ to the future without any knowledge of the parameters of the intermediate expansion of the universe. Thus, in order to communicate the sender has to prepare the two fields in a state $F_{\lambda,k}$.

{\it--- Example b: accelerated observer.}
Consider a communication protocol between an inertial and uniformly accelerated observers in flat space-time. Suppose that an inertial observer wishes to send a classical number $\lambda$ to a uniformly accelerated recipient moving with unknown proper acceleration. In order to do that the sender has to prepare the eigenstate \eqref{eigenfunction_2} and the accelerated receiver has to measure the operator $\hat{L}$ leading to the retrieval of the encoded number $\lambda$. The corresponding Bogolyubov transformation between the modes and operators in Minkowski (inertial) and Rindler (noninertial) frame of reference involves mixing all the frequencies \cite{Bruschi2010}. The relation between the operators in Minkowski space ($\hat{a}_l$) and Rindler space ($\hat{b}_{k,\text{I}},\,\hat{b}_{k,\text{II}}$) is given by:
\begin{align}
\hat{b}_{k,\text{I}}&=\int \text{d}l\, \alpha_{kl}^*(\hat{a}_l+e^{-\pi c^2|k|/a}\hat{a}_l^\dagger),\\
\hat{b}_{k,\text{II}}&=\int \text{d}p\, \alpha_{kp}(\hat{a}_p+e^{-\pi c^2|k|/a}\hat{a}_p^\dagger).
\end{align}
where $\text{I,II}$ refer to the Rindler wedges, $a$ is the (uniform) acceleration of the non-inertial observer and:
\begin{align}
\alpha_{kl}&=\frac{k}{4\pi a\sqrt{|kl|}}\left(\frac{a}{il}\right)^{ik/a}\left(\frac{k}{|k|}+\frac{l}{|l|}\right)\Gamma\left(ik/a\right);
\end{align}
with the analogous relation for $\tilde{\phi}$. 
One needs to ask whether it is enough to be localized in only one region to successfully encrypt the information. 
Using the properties of the Bogolyubov coefficients $\alpha_{kl}$ and corresponding canonical constraints it can be shown that one can represent operator $\hat{L}$ as sum of operators $\hat{L}_\text{I}$ and $\hat{L}_{\text{II}}$ acting in region $I$ and $II$ exclusively:
\begin{align}
\hat{L}&=\int \text{d}k \hat{L}_k=\hat{L}_\text{I}+\hat{L}_{\text{II}}=\int \text{d}p \hat{L}_{p,\text{I}}+\int \text{d}s \hat{L}_{s,\text{II}}=\nonumber\\
=&-i\int \text{d}p (\hat{b}^\dagger_{p,\text{I}}\hat{\tilde{b}}_{p,\text{I}}-\hat{b}_{p,\text{I}}\hat{\tilde{b}}^\dagger_{p,\text{I}})+\nonumber\\
&-i\int \text{d}s (\hat{b}^\dagger_{s,\text{II}}\hat{\tilde{b}}_{s,\text{II}}-\hat{b}_{s,\text{II}}\hat{\tilde{b}}^\dagger_{s,\text{II}}).\label{Eq21}
\end{align}
Therefore, 
In the following we show that previously discussed states:
\begin{align}
F_{\lambda}(\{x,\tilde{x}\})&=\Pi_k f_{\lambda,k}(x_k,\tilde{x}_k) g_{k}(x^2_k+\tilde{x}^2_k),
\end{align}
are also eigenstates of $\hat{L}_\text{I}$ and $\hat{L}_{\text{II}}$ alone. 
Without loss of generality we choose a specific function $g_{k}(x^2_k+\tilde{x}^2_k)=e^{-x_k^2-\tilde{x}_k^2}$ and evaluate $\hat{L}_\text{I}F_{\lambda}(\{x,\tilde{x}\})$ which produces:
\begin{align}
&\hat{L}_\text{I}F_{\lambda}(\{x,\tilde{x}\})=-\frac{i}{2}\int \text{d}k\text{d}p\text{d}l\bigg(w_{klp}x_l\tilde{x}_p+\nonumber\\
&+2z_{klp}x_l\tilde{x}_p\big(\sum_s(\delta_{ps}-\delta_{ls})\big)F_{\lambda}(\{x,\tilde{x}\})+\nonumber\\
&+i\lambda z_{klp}\big(\sum_s\delta_{ls}\frac{\tilde{x}_l\tilde{x}_p}{x_l^2+\tilde{x}_p^2}+\delta_{ps}\frac{x_lx_p}{x_p^2+\tilde{x}_l^2}\big)F_{\lambda}(\{x,\tilde{x}\})\bigg),
\end{align}
where:
\begin{align}
w_{klp}&=2e^{-\pi c^2|k|/a}(1+\cosh[\pi c^2|k|/a])\big(\alpha_{kl}\alpha_{kp}^*-\alpha_{kl}^*\alpha_{kp}\big),\nonumber\\
z_{klp}&=(1-e^{-2\pi c^2|k|/a})\big(\alpha_{kl}\alpha_{kp}^*+\alpha_{kl}^*\alpha_{kp}\big).
\end{align}
By means of the canonical properties of the inverse Bogolyubov transformation, which give $\int \text{d}k(\alpha_{kl}^*\alpha_{kp}-\alpha_{kl}\alpha_{kp}^*e^{-2\pi c^2|k|/a})=\delta_{pl}$ and $\int \text{d}k e^{-\pi c^2|k|/a}(\alpha_{kl}^*\alpha_{kp}-\alpha_{kl}\alpha_{kp}^*)=0$, one can arrive at the eigenequation of the form:
\begin{align}
\hat{L}_\text{I}F_{\lambda}(\{x,\tilde{x}\}) = \lambda\int \text{d}p\,1 F_{\lambda}(\{x,\tilde{x}\}).\label{accelerated-eigenequation}
\end{align}
To avoid any divergences and in analogy to what have been done previously, one would like to introduce in the definition of $\hat{L}$ and $\hat{L}_I$ a normalization function $\rho(k)$. This would lead to modifications of canonical constraints (of the form $\int \text{d}k \rho(k)\alpha_{lk}\alpha_{pk}^*$ and $\int \text{d}k\rho(k)(\alpha_{kl}^*\alpha_{kp}-\alpha_{kl}\alpha_{kp}^*e^{-2\pi c^2|k|/a})$) and equations Eq.(\ref{Eq21}) and Eq.(\ref{accelerated-eigenequation}). From the physical point of view this would correspond to measurements of a limited window of frequencies, thus yielding a finite result. Any measurement of such form will inevitably induce error in evaluating $\lambda$. The latter scenario will be the subject of further study as it must involve a broader analysis of the suggested communication protocol and spatial localization of the detectors as in \cite{Dragan2013a,Dragan2013b}.

{\it ---Conclusions.} 
We have shown how two observers without a shared reference frame can communicate using quantum fields in relativistic settings. The unspecified Bogolyubov transformation between the respective frames changes the fields, however certain correlations between different fields are preserved. We encode the information in the correlated states to protect it from the influence of the unknown transformation.

The reason why reliable communication protocol can be introduced is the symmetry of the transformation applied to the transmitted states. In our case it is the fields' interchange symmetry of the Hamiltonian \eqref{hamiltonian}. However it should be expected that any other type of transformation symmetry can be used to send information across. For example, if the transformation is symmetric under time translation, one can use temporal correlations as carriers of information, as described in \cite{Ball2004}. An analogous protocol would also apply in the case of spatial translation symmetries. In general, any type of symmetry leads to preservation of certain correlations. Therefore one can expect an interesting relation between Noether's theorem linking symmetries of the dynamics and preserved currents, with fundamental ability to communicate in the presence of the dynamics. This is currently a subject of our further investigation.

The results are applicable not only to the case of relative motions of the observers but also any other physical settings, where quadratic Hamiltonians or Bogolyubov transformations play a role.

\section*{Acknowledgements}

Authors would like to thank Konrad Banaszek, Pieter Kok and David Edward Bruschi for useful discussions. We acknowledge financial support from the National Science Center, Sonata BIS Grant No. 2012/07/E/ST2/01402.

\end{document}